\documentclass[letterpaper]{JHEP3}
\usepackage{epsfig,amsmath,xspace}

\DeclareMathOperator{\Tr}{Tr}

\newcommand{\rh}{r_{\rm h}}
\newcommand{\ts}{\tilde\sigma}

\newcommand{\GN}{G_{\rm N}}
\newcommand{\R}{\mathbf{R}}
\newcommand{\Z}{\mathbf{Z}}

\title{Hedgehog black holes and the Polyakov loop at strong coupling}

\author{
Matthew Headrick \\ 
Stanford Institute for Theoretical Physics, Stanford CA 94305-4060, USA \\ 
\email{headrick@stanford.edu}
}

\abstract{
In $\mathcal{N}=4$ super-Yang-Mills theory at large $N$, large $\lambda$, and finite temperature, the value of the Wilson-Maldacena loop wrapping the Euclidean time circle (the Polyakov-Maldacena loop, or PML) is computed by the area of a certain minimal surface in the dual supergravity background. This prescription can be used to calculate the free energy as a function of the PML (averaged over the spatial coordinates), by introducing into the bulk action a Lagrange multiplier term that fixes the (average) area of the appropriate minimal surface. This term, which can also be viewed as a chemical potential for the PML, contributes to the bulk stress tensor like a string stretching from the horizon to the boundary (smeared over the angular directions). We find the corresponding ``hedgehog" black hole solutions numerically, within an $SO(6)$-preserving ansatz, and derive part of the free energy diagram for the PML. As a warm-up problem, we also find exact solutions for hedgehog black holes in pure gravity, and derive the free energy and phase diagrams for that system.
}

\preprint{SU-ITP-07/23}

\begin{document}

\section{Introduction}

There has recently been a lot of interest in the large-$N$ thermodynamics of four-dimensional $SU(N)$ gauge theories compactified on $S^3$. Although they live on a space with finite volume, in the 't Hooft limit such theories have an infinite number of degrees of freedom and can therefore exhibit sharp phase transitions and spontaneous symmetry breaking. Much of the recent interest stems from the observation by Sundborg \cite{Sundborg:1999ue} and Aharony et al.\ \cite{Minwalla} that, even in the free limit, such theories can undergo a first-order phase transition as a function of temperature, and this transition shares enough features of the usual deconfinement transition in flat space to deserve the same name. Specifically, the free energy goes from being of order $N^0$ to $N^2$, indicating liberation of the elementary colored fields, and the Polyakov loop (the Wilson loop wrapping the periodic imaginary time direction) becomes non-zero, indicating breaking of the $\Z_N$ center symmetry. The strategy employed in that work was to compute the effective potential for a single mode of the system, namely the constant mode of the Polyakov loop over the three-sphere, by integrating out all other modes. From this effective potential, or off-shell free energy, one can read off the saddle points of the theory---stable, unstable, and meta-stable---and how they depend on temperature. Much of the subsequent work on weakly-coupled large $N$ gauge theories on $S^3$ has continued to focus on the Polyakov loop and its effective potential \cite{Liu:2004vy,Aharony:2005bq,Yamada:2006rx,Hollowood:2006xb,Hartnoll:2006pj,Unsal:2007fb,Gursoy:2007np,Aharony:2007rj,Dutta:2007ws,Hoyos:2007ds}.

At large 't Hooft coupling, the easiest theory in this class to study is the conformal $\mathcal{N}=4$ super-Yang-Mills theory, which is dual to type IIB supergravity on asymptotically AdS${}_5\times S^5$ manifolds. At finite temperature, gravitational systems with a negative cosmological constant and asymptotically AdS boundary conditions are subject to a Hawking-Page phase transition \cite{Hawking:1982dh}. This is a first-order transition, below which the dominant thermal state is AdS and above which it is a large black hole. (We briefly review this transition in Section 2.) From the gauge theory point of view, this is again a deconfinement transition \cite{Witten:1998zw}, with the same two properties mentioned above: the free energy goes from being of of order $N^0$ to $N^2$, and the $\Z_N$ center symmetry breaks. The former effect is due to the fact that the black hole solution has a non-zero classical action, and this action carries an overall factor of $1/G_{\rm N}\sim N^2$. The latter effect can be seen as follows. While the usual Wilson loop is not straightforward to compute using AdS/CFT (but see \cite{Alday:2007he} for a proposal), one can compute a locally BPS version that includes a coupling to the six adjoint scalars $\phi^I$:
\begin{equation}\label{PMLdef}
W(C) = \frac1N\Tr M(C)\,, \qquad M(C) = P\exp\oint_Cds(iv^\mu A_\mu + |v|\tilde\theta^I\phi^I)\,.
\end{equation}
(Here $v^\mu=dx^\mu/ds$ along the curve $C$, and $\tilde\theta^I$ is an arbitrary unit six-vector.) This is the so-called Wilson-Maldacena loop, and it is given as $e^{-s_{\rm dominant}}$, where $s_{\rm dominant}$ is the saddle-point value of the action of a string world-sheet anchored on the curve $C$ \cite{Maldacena:1998im,Rey:1998ik}. In the absence of a $B$-field, $s_{\rm dominant}$ is simply the area in string units of the minimal surface with those boundary conditions. (More precisely, this area is divergent due to the region near the boundary---a UV divergence from the boundary point of view---and must be renormalized.) Just like the Polyakov loop, the Wilson-Maldacena loop running around the imaginary time circle (the so-called Polyakov-Maldacena loop, or PML) is charged under the $\Z_N$ center. In the Euclidean version of AdS the imaginary time circle is non-contractible, so there actually aren't any world-sheets with the appropriate boundary conditions, implying that the PML vanishes and the center is unbroken. On the other hand, due to the existence of a horizon the imaginary time circle for a Euclidean black hole is contractible, leading to a finite value for $s_{\rm dominant}$ and hence the PML. The Lorentzian version of these statements is that the string ending on the horizon represents screening of the quark charge in the field theory; in the absence of a horizon the string has nowhere to end, and therefore inserting a single quark is forbidden.\footnote{The discussion in this paragraph is strictly speaking valid only at infinite $N$. At finite $N$ the system has a finite number of degrees of freedom and therefore no spontaneous symmetry breaking or phase transitions are possible. Correspondingly, the PML vanishes identically due to the Gauss law constraint. On the AdS side this is enforced by the integral over a certain $B$-field Wilson line \cite{Witten:1998zw}. In order to see the effects described in this paragraph, therefore, it is necessary to take the large $N$ limit in the presence of a fixed but very small explicit breaking of the center. See \cite{Minwalla} for a detailed discussion of this point. In Section 3, we will define our order parameter in such a way that this issue does not arise, by also adding an anti-quark at the antipodal point on the $S^3$.}

Much of the subsequent work on the thermodynamics of this system has focused on either the PML or the ordinary Polyakov loop, and their respective effective potentials \cite{Barbon:2001di,Barbon:2002nw,Barbon:2004dd,AlvarezGaume:2005fv,Kruczenski:2005pj,Hartnoll:2006hr,AlvarezGaume:2006jg,Azuma:2007fj}. In the papers \cite{AlvarezGaume:2005fv,AlvarezGaume:2006jg,Azuma:2007fj}, Wadia and collaborators studied effective potentials for the ordinary Polyakov loop based on phenomenological unitary matrix models. On the other hand, conjectural sketches of the free energy diagram for the PML, built around known properties of the saddle points, have appeared in various places, including \cite{Barbon:2001di,Barbon:2002nw,Barbon:2004dd,Kruczenski:2005pj,kumarnaqvi}. Our purpose in this paper is to compute quantitatively the effective potential for the PML, using the gravitational dual theory. As we will now briefly explain, this will lead us to a problem in General Relativity that is of some interest in its own right. Due to the complexity of the gravitational theory, we will in this first pass only be partially successful, and part of the free energy diagram will remain a conjectural sketch. Based on our results, we will also sketch a phase diagram for the system in the presence of a chemical potential for the PML (roughly speaking a density of external quarks).

The basic method we will use to compute the free energy diagram for the PML is due to York \cite{York:1986it}; let us briefly review his work. He was studying gravity in a cavity at finite temperature (the thermodynamics of this system is closely analogous to that of asymptotically AdS spaces; the calculation described below was done in the asymptotically AdS case in \cite{Barbon:2001di,Barbon:2002nw,Barbon:2004dd}). He considered sub-ensembles of the canonical ensemble in which the black hole horizon area was fixed. To do this, he added a Lagrange multiplier term to the Euclidean Einstein-Hilbert action. The saddle-point approximation to the free energy is then given by the action evaluated on the solution to the new equations of motion. The effect of the Lagrange multiplier term is to add to the stress tensor in the Einstein equation a term corresponding to a membrane wrapping the horizon. This stress-energy  creates a conical deficit angle in the orthogonal (i.e.\ imaginary time--radial) plane. The result is a free energy diagram that clearly shows the main features of the system's thermodynamics: the three saddle points, namely hot flat space and large and small black holes, the thermodynamic instability of the small black hole, and the Hawking-Page transition (see figure 11 of \cite{Headrick:2006ti}).
Interestingly, the free energy turns out to be continuous even across the topology-changing transition separating the flat space and black hole regimes.

The idea of applying the same method to computing the free energy diagram for the PML  was discussed by Wiseman and the present author in the paper \cite{Headrick:2006ti}. As above we should add a Lagrange multiplier term to the bulk gravity action, but this time fixing the area not of the horizon but of the orthogonal disc, spanned by the radial and imaginary time directions. Its effect is to add to the stress tensor a term corresponding to a fundamental string stretching from the horizon to the boundary. 
More precisely, to define the PML one must fix a point on the boundary $S^3$ as well one as on the asymptotic $S^5$. To have the lightest possible mode as our order parameter, we take its (logarithmic) average over both spheres. (A precise definition of our order parameter is given in Section 3. It is morally similar to the one used by Aharony et al. \cite{Minwalla,Aharony:2005bq}, but here the averaging is done in a manifestly gauge-invariant, rather than gauge-fixed, manner.) We thus avoid explicitly breaking the system's $SO(4)\times SO(6)$ symmetry. Assuming that this symmetry is not spontaneously broken (which it is in some cases, as we will discuss below), the result is to smear the string source uniformly over both sets of angular directions.

We are thus left with the problem of finding the solution to the Einstein equation for a black hole with a relativistic string stretching from the horizon to infinity and smeared over the angular directions. That problem was solved for pure gravity in four dimensions by Guendelman and Rabinowitz \cite{Guendelman:1991qb}, 
who termed the solutions ``hedgehog black holes". As a warm-up problem, in Section 4 we generalize their solutions to arbitrary dimensions. The solutions are beautifully simple, as are the free energy and phase diagrams derived from them (see figures \ref{kappaFpureS3} and \ref{phasepureS3}).
As in York's case, the free energy is continuous even as the topology changes from the black hole to AdS.
Perhaps the most interest feature of the phase diagram is that the Hawking-Page transition curve ends at a critical point.

The actual system of interest, supergravity on AdS${}_5\times S^5$, is discussed in Section 5. Thankfully, the symmetries of the problem remove most of the new fields; the only ones remaining are the dilaton and a scalar representing the radius of the five-sphere. Thus, after gauge fixing, there are four degrees of freedom (the dilaton and the radii of the $S^1$, $S^3$, and $S^5$) depending on one variable. Nonetheless the equations of motion are sufficiently complicated that they defy exact solution, and we had to resort to solving them numerically. The resulting free energy diagram is shown in figure \ref{FS3}. One unexpected feature is the existence of a (temperature-dependent) positive lower bound on the logarithm of the PML. However, it should be recalled that we have worked within an ansatz that preserves the $SO(6)$ symmetry of the boundary conditions. It is known that, at sufficiently high temperature, the conventional small black hole with that symmetry is unstable to perturbations that break it \cite{Hubeny:2002xn}. It is also known that there exists at least one solution with lower free energy in which $SO(6)$ is spontaneously broken, namely the ten-dimensional black hole \cite{Horowitz:2000kx}. This conventional solution should extend into a branch of ten-dimensional hedgehog solutions, which we conjecture continues all the way to unit PML. Finding these solutions, however, would present a challenging numerical problem, and no attempt is made to do so here (even the conventional ten-dimensional black hole is not known analytically or numerically, outside of the limit that its horizon is much smaller than the AdS radius \cite{Horowitz:2000kx}). In the phase diagram for the system with a chemical potential for the PML, the Hawking-Page transition should extend to a first-order transition separating five- and ten-dimensional black hole phases. We sketch a possible such phase diagram in figure \ref{phaseS3}.

For completeness we have also done the analogous calculations for the gauge theory on $\R^3$, which can be considered as the high-temperature limit of the theory on $S^3$ (subsection 5.2).

Let us close this introduction with two general comments. First, one might ask what happens if one retains the entire matrix $M$ \eqref{PMLdef} whose trace is the PML, rather than integrating out everything but the PML itself. In many instances such matrix models capture the interesting physics of a system more directly than is revealed by computing the effective potential for the trace alone. Finding the matrix model governing $M$ is equivalent to computing the effective action for the collection of all possible traces,
\begin{equation}\label{otherWs}
W_{n_1n_2n_3\cdots} = \frac1N\Tr(M^{n_1}M^{\dag n_2}M^{n_3}\cdots)\,,
\end{equation}
where $(n_1,n_2,n_3,\dots)$ is an arbitrary collection of (positive or negative) integers (recall that $M$ is a general matrix, neither Hermitian nor unitary). Unfortunately, except for the simple traces of the form $W_n = \Tr M^n/N$ and $\Tr M^{\dag n}$, it is not clear how to compute such observables using AdS/CFT. Furthermore, even those simple traces are not independent observables in the supergravity limit, since they obey the identity $W_n = W^{|n|}$ (up to $\alpha'$ and $1/N$ corrections), since the minimal surface for the $n$-fold cover of a curve is simply the $n$-fold cover of the basic minimal surface (putting aside the special case where the fundamental group of the bulk topology has a finite part). Thus it is not possible, by adjusting the supergravity configuration, to independently fix $W_n$ for different $n$. To summarize, in the supergravity limit the effective potential for $W$ is ``all there is".

A second comment is that, in the papers \cite{Barbon:2001di,Barbon:2002nw,Barbon:2004dd,AlvarezGaume:2005fv,Kruczenski:2005pj,AlvarezGaume:2006jg}, one of the motivations for going off-shell was to study the Hagedorn transition of string theory, which on-shell is hidden behind the Hawking-Page transition. In particular these papers conjecture that the Hagedorn transition occurs when the thermal AdS and small black hole (or possibly ten-dimensional black hole) saddle points merge. Unfortunately, since we work within the gravity approximation, our free energy diagram is too crude to address this stringy issue. However, as a matter of principle our general method for going off-shell should be applicable to the full string theory; it would be interesting to see whether it can be applied in practice.

\section{Review: Hawking-Page transition}

The purpose of this section is to remind the reader of the basic facts about the Hawking-Page transition \cite{Hawking:1982dh}, while setting up the framework in which we work.

We consider four-dimensional $SU(N)$ $\mathcal{N}=4$ super-Yang-Mills theory, compactified on a unit three-sphere. In the regime of parameters $1\ll\lambda\ll N^2$, where $\lambda=g_{\rm YM}^2N$ is the 't Hooft coupling, the theory is best described by type IIB supergravity (UV completed by type IIB string theory) on asymptotically AdS${}_5\times S^5$ manifolds. The curvature is supported by $N$ units of five-form flux on the $S^5$. Stringy effects are suppressed by $1/\lambda$ and quantum effects by $1/N$; more precisely, setting the $S^5$ and AdS${}_5$ radii to one, the ten-dimensional gravitational coupling constant is $\GN = \frac12\pi^4N^{-2}$ while the string length is $\alpha^{\prime1/2} = \lambda^{-1/4}$. 

The partition function for the canonical ensemble at temperature $T$ is defined by the Euclidean path integral on $S^1\times S^3$, where the imaginary time circle has circumference $1/T$ and anti-periodic boundary conditions for the fermions. It is computed by the gravitational path integral over Euclidean manifolds that are asymptotically AdS${}_5^T\times S^5$, where AdS${}_5^T$ is Euclidean AdS${}_5$ with ``imaginary time" coordinate periodically identified so that its conformal boundary is the same $S^1\times S^3$ on which the field theory lives. Again, at large $N$ the saddle point approximation is valid, and we have (to lowest order in $1/N$)
\begin{equation}
e^{-F/T} = Z \approx e^{-S_{\rm dominant}}\,,
\end{equation}
where $S_{\rm dominant}$ is the minimum value of the Euclidean action among the saddle points. Actually, given the boundary conditions, on any solution this action will be divergent due to the infinite volume near the boundary (an ultraviolet divergence from the field theory viewpoint), and must therefore be regularized\footnote{\label{regularization}The simplest regularization procedure is to impose a finite Dirichlet boundary condition on the metric. We require the induced metric to be $S^1_{R/(2\pi T)}\times S^3_R\times S^5_1$, where the subscripts indicate the radii of the respective spheres. $R$ is taken to infinity to remove the regulator.} and made finite by a divergent subtraction; conventionally the counterterm is chosen so that AdS${}_5^T\times S^5$ has vanishing renormalized action. Since the action comes multiplied by an overall $1/\GN\sim N^2$, the free energy is of order $N^2$, unless the action of the dominant saddle point vanishes. In the latter case the one-loop contribution, which is of order $N^0$, is the leading term; it represents the contribution from the thermal gas of gravitons.

At any temperature, AdS${}_5^T\times S^5$ is a saddle point, and at temperatures below $T_{\rm crit} = 8^{1/2}/(2\pi)$ it is the only one. Above $T_{\rm crit}$ there are at least two others, namely the small and large AdS-Schwarzschild black holes. Both solutions take the form of a product of $S^5$ with the following asymptotically AdS${}_5^T$ Einstein metric:
\begin{gather}\label{BHmetric}
ds^2 = f(r)d\tau^2 + f(r)^{-1}dr^2 + r^2d\Omega_3^2\,,\\
f(r) = r^2 + 1 - \frac\mu{r^2}\,, \qquad \tau\sim\tau+\frac1T, \qquad \rh<r<\infty\,,
\end{gather}
where the horizon radius $\rh$ is the largest zero of $f(r)$, and $\mu$ is fixed by the requirement of smoothness at the horizon, $f'(\rh)=4\pi T$. None of the supergravity fields aside from the metric and five-form are excited. Eliminating $\mu$ from the two equations $f(\rh) = 0$ and $f'(\rh)=4\pi T$, we arrive at a quadratic equation for $\rh$. For $T>T_{\rm crit}$, this equation admits two solutions, one with $\rh<2^{-1/2}$, the other with $\rh>2^{-1/2}$; these are the small and large black holes respectively. Their actions are
\begin{equation}
S = \frac{N^2}{4T}\rh^2(1-\rh^2)\,.
\end{equation}
Any black hole with $\rh>1$ is therefore thermodynamically dominant over thermal AdS; this is never the case for the small black hole, but is the case for the large black hole when $T>T_{\rm HP}\equiv 3/(2\pi)$.

Besides never being thermodynamically dominant, the small black hole has a negative eigenmode of the Lichnerowicz operator \cite{Prestidge:1999uq} (similar to the Schwarzschild solution's Gross-Perry-Yaffe mode \cite{Gross:1982cv}). This is related to the fact that its Lorentzian version represents a black hole in an unstable thermal equilibrium with the heat bath, due to its negative specific heat. Furthermore, for $\rh$ less than around $0.426$, which translates to temperatures above around $3.20/(2\pi)$ (slightly higher than $T_{\rm HP}$), the small black hole suffers from a second, dynamical instability, namely a Gregory-Laflamme instability on the $S^5$ \cite{Hubeny:2002xn}.

The large and small AdS-Schwarzschild black holes are uniform on the $S^5$, and therefore preserve the boundary conditions' $SO(6)$ symmetry. At sufficiently high temperatures, there are other saddle points that spontaneously break that symmetry. These presumably include a black hole that is spread out non-uniformly over the $S^5$, and a ten-dimensional black hole \cite{Horowitz:2000kx} that is localized on the $S^5$ and breaks the $SO(6)$ down to $SO(5)$; its horizon topology is $S^8$, rather than $S^3\times S^5$ in the case of the AdS-Schwarzschild black holes. At high temperatures, where its horizon is very small compared to the AdS radius, this is essentially a ten-dimensional Schwarzschild black hole, and is therefore presumably dynamically stable but thermodynamically unstable. Exact solutions for these saddle points (or others with more exotic horizon topologies) are still unknown, as are their precise properties, such as at what temperatures they appear.

\section{Order parameter and off-shell free energy}

The Hawking-Page phase transition in AdS${}_5\times S^5$ described in the previous section is, from the point of view of the boundary gauge theory on $S^3$, a deconfinement transition \cite{Witten:1998zw}. There are two features which qualify it for this name. First, as explained above, the free energy goes from being of order $N^0$ to $N^2$. Second, the $\Z_N$ center symmetry breaks. 
In this section we will study this breaking by defining a certain order parameter for the symmetry. We will then explain how, in the rest of the paper, we will compute the off-shell free energy as a function of that order parameter.

We start by defining the Polyakov-Maldacena loop (PML). This is a function on $S^3\times S^5$. Using $\tau$ and $\theta$ to denote the coordinates on the $S^1$ and $S^3$ respectively, and $\tilde\theta^I$ to denote a unit vector in $\R^6$, we have \cite{Maldacena:1998im,Rey:1998ik}
\begin{equation}
W(\theta,\tilde\theta) \equiv
\frac1N\Tr P\exp\oint d\tau
\left(iA_\tau(\tau,\theta) + \tilde\theta^I\phi^I(\tau,\theta)\right).
\end{equation}
Here $A$ is the gauge field and $\phi^I$ are the six adjoint scalars. $W$ has unit charge under the $\Z_N$ center symmetry; it could therefore serve as an order parameter, except that, for the very same reason, its expectation value vanishes identically (no spontaneous symmetry breaking can occur in a system with a finite number of degrees of freedom\footnote{An equivalent way to see that its expectation value vanishes is to use the Gauss law constraint; the charge of a single external quark cannot be neutralized by the elementary quanta of the theory, which all transform in the adjoint representation, to produce a singlet under the global part of the gauge group.}) \cite{Witten:1998zw}. Therefore we multiply $W$ by the PML $W^\dag$ for an antiquark, which has charge $-1$ under the $\Z_N$, that we choose to insert at the same point on the $S^5$ but at the antipodal point $-\theta$ on the $S^3$:
\begin{equation}
w(\theta,\tilde\theta) \equiv W(\theta,\tilde\theta)W^\dag(-\theta,\tilde\theta)\,.
\end{equation}
Although neutral under the $\Z_N$ center of the gauge group, $w$ nonetheless serves as an order parameter; its expectation value is of order $N^{-2}$ when the symmetry is unbroken, and otherwise of order $N^0$.\footnote{This fact is explained in detail in \cite{Minwalla}. It can be understood roughly from the fact that large $N$ factorization, which would have implied $\langle w\rangle = \langle W\rangle \langle W^\dag\rangle+\mathcal{O}(N^{-2})$, fails only because in the path integral we must sum over gauge configurations that are related by the action of the center symmetry; this sum sets $\langle W\rangle$ to zero but not $\langle w\rangle$. Aside from this subtlety, the operators we deal with become classical in the large $N$ limit, so we will generally simply refer to their ``values" rather than their ``expectation values".} We have
\begin{equation}
w = e^{-f_{q\bar q}/T}\,,
\end{equation}
where $f_{q\bar q}$ is the free energy cost of adding to the system an external quark and anti-quark (charged under the scalars as well as the gauge field). Because of the self-energies of the quark and anti-quark, $f_{q\bar q}$ suffers from an ultraviolet divergence and must be renormalized.

In the gravity dual, the external quark and anti-quark source fundamental strings that extend into the bulk (with opposite orientations, i.e.\ if we put an arrow on the string, then the arrow would point away from the quark and towards the anti-quark). If we are at large $N$, then as in the previous section we appeal to the saddle-point approximation and imagine that the bulk is described by a fixed classical supergravity solution. If we are also at large $\lambda$, then the quantum fluctuations of the world-sheet are also suppressed, and to a good approximation the string world-sheet lies on a fixed surface in the bulk geometry, namely the solution to its classical equations of motion (if there are several solutions, then the one with the smallest Euclidean action). The quark and anti-quark world-lines on the boundary provide the boundary conditions for this surface, along with the value of $\tilde\theta$, which fixes the asymptotic position of the string on the $S^5$. We have \cite{Maldacena:1998im,Rey:1998ik}:
\begin{equation}
w = \frac1{N^2}e^{-s_{\rm dominant}}\,,
\end{equation}
where we use $s$ for the Euclidean world-sheet action. (The factor of $N^{-2}$ arises from the same factor in the definition of $w$.) $s_{\rm dominant}$ is divergent due to the infinite area of the world-sheet near the boundary (reflecting the ultraviolet divergence of $f_{q\bar q}$), and must be renormalized, similarly to the bulk action $S$.

The world-sheet action $s$ is, for our purposes, made up of two terms. The topological (Fradkin-Tseytlin) term contributes a factor of $N^2$ to $e^{-s}$ if the world-sheet is topologically two discs, and $N^0$ if it is a cylinder (just as in the usual 't Hooft genus counting). Therefore a saddle point with two discs, if it exists, will always dominate in 't Hooft limit over a cylinder. The Nambu-Goto term, which is the (string-frame) area of the world-sheet in string units, requires the world-sheet to be the minimal surface within each topological sector. (If the supergravity background has a $B$-field, then it also needs to be taken into account in minimizing the world-sheet action. Non-zero Ramond-Ramond fields---which are of course present in the case at hand---couple only to the world-sheet fermions and therefore should not affect the classical solutions.)

Let us first consider the case where the bulk metric is the Euclidean black hole solution \eqref{BHmetric}. The dominant world-sheet configuration has two connected components, which wrap the disc and are constant on the $S^3$ at the points $\theta$ and $-\theta$ respectively, and also constant on the $S^5$ at the point $\tilde\theta$. The area of the disc is easily read off from the metric \eqref{BHmetric} \cite{Kruczenski:2005pj}:
\begin{equation}
A_{\rm ct} + \frac1T\int_{\rh}^\infty dr = -\frac{\rh}T\,,
\end{equation}
where we have added an $\rh$-independent counterterm ($-R/T$ in the regularization described in footnote \ref{regularization}). We thus find
\begin{equation}
f_{q\bar q} = -\frac{2\rh}{2\pi\alpha'} = -\frac{\lambda^{1/2}\rh}{\pi}\,.
\end{equation}

On the other hand, when the bulk metric is thermal AdS, the Euclidean time-circle is non-contractible, so there are no world-sheet configurations that are topologically two discs. The dominant one is instead a cylinder, with the string connecting the quark to the anti-quark through the center of AdS, and wrapping the $S^1$ (again at a fixed position on the $S^5$). This world-sheet has vanishing renormalized area, so we have
\begin{equation}\label{wAdS}
w = \frac1{N^2}\,.
\end{equation}

In the rest of the paper, we will be interested in fixing a particular value for the PML, and integrating out all other degrees of freedom. More specifically, we will take as our order parameter the average value of $f_{q\bar q}$ over the three- and five-spheres (times $-\pi\lambda^{-1/2}$):\footnote{It is interesting to compare our order parameter with the one employed by Aharony et al. \cite{Minwalla} for their weak-coupling calculations:
\begin{equation}\label{theirOP}
\frac1N\Tr P\exp\oint d\tau\,i\int\frac{d^3\theta}{\Omega_3}A_\tau^{\rm gauge-fixed}(\tau,\theta)\,,
\end{equation}
where they have picked a particular gauge. Gauge-fixing is necessary for them in order to make the average of $A_\tau$ over the three-sphere a sensible (i.e.\ gauge-invariant) observable. Thus one difference between our order parameter and theirs is that we take this average after computing $\ln N^{-1}\Tr P\exp\oint d\tau$, whereas they take it before. Another difference is that they do not include a coupling to the scalars (therefore they have no need to average over $S^5$).
}
\begin{equation}\label{OPdef}
\Phi \equiv \frac{\pi T}{\lambda^{1/2}}\int\frac{d^3\theta}{\Omega_3}\frac{d^5\tilde\theta}{\Omega_5}\ln w\,.
\end{equation}
The reason for taking the average over the spheres is to retain the lightest possible mode when we integrate out the other modes. In this way, we avoid explicitly breaking the theory's $SO(4)\times SO(6)$ global symmetry. The factor of $\lambda^{-1/2}$ is for convenience, in order to have a quantity which is finite in the limit $\lambda\to\infty$. We have, for example, $\Phi = \rh$ for AdS-Schwarzschild and $\Phi = -2\pi T\lambda^{-1/2}\ln N$ for pure AdS (the latter value goes to $-\infty$ in the strict 't Hooft limit; on the other hand it goes to zero if we take $N$ to infinity holding, say, $g_{\rm YM}$ fixed, which is also well within the regime of validity of supergravity). Those solutions both have $SO(4)\times SO(6)$ isometries which dictate the position of the classical string world-sheet, and therefore $\Phi$ reduces to a local functional of the metric; we will lean heavily on this convenient property in our computations below. In general, without those isometries, $\Phi$ is a non-local functional of the metric since one must solve the string equations of motion.

We now define the effective potential (or off-shell free energy) $F(\Phi)$ by
\begin{equation}\label{Fdef}
e^{-F(\Phi_0)/T} = \int[d\phi]\delta(\Phi[\phi]-\Phi_0)e^{-S[\phi]}\,.
\end{equation}
Here $\phi$ includes all the degrees of freedom of the system---in our case, all modes of all the supergravity fields. We then have
\begin{equation}
e^{-F/T} = Z = \int d\Phi\,e^{-F(\Phi)/T}\,.
\end{equation}
In the thermodynamic limit (in our case, the large $N$ limit) this last integral is evaluated using the saddle point method, i.e.\ simply by finding the global minimum of $F(\Phi)$, which represents the thermal equilibrium; other local minima represent metastable equilibrium configurations.

Our interest will be in evaluating the right-hand side of \eqref{Fdef}, which we also will do by a saddle-point method. In order to restrict the integral to configurations with the given value of $\Phi[\phi]$, we add to the supergravity action a Lagrange multiplier term,
\begin{equation}
S_{\rm SUGRA}[\phi] \to 
S_{\rm SUGRA}[\phi] - \frac\kappa T\left(\Phi[\phi]-\Phi_0\right).
\end{equation}
We then find the solutions to this modified action, and evaluate the action on them. The effect of the new term on the equations of motion will be discussed in detail in the next two sections.

\section{Warm-up: pure gravity}

In this section we will solve the problem explained in the last section for a simpler theory than type IIB supergravity, namely pure gravity with negative cosmological constant. We will see that this system is simple enough to allow an exact solution. 

It turns out to be just as easy to work in a general spacetime dimension $D=p+2$ ($p\ge1$). As above we set the AdS radius to one. Again, we work in the canonical ensemble at a temperature $T$, so we consider Euclidean metrics that are asymptotically AdS${}_{p+2}^T$, that is, whose conformal boundary is $S^1_{1/(2\pi T)}\times S^p_1$. These boundary conditions have an $SO(2)\times SO(p+1)$ symmetry, which we assume is shared by the metric; in other words we look for static, spherically symmetric solutions. We thus take the following metric ansatz:
\begin{equation}
ds^2 = g_{\tau\tau}(r)d\tau^2 + g_{rr}(r)dr^2 + g_{\Omega\Omega}(r)d\Omega_p^2\,,\qquad
\tau\sim\tau+\frac1T\,, \qquad \lim_{r\to\infty}\frac{g_{\tau\tau}}{g_{\Omega\Omega}} = 1\,.
\end{equation}
In this context, our order parameter \eqref{OPdef} becomes
\begin{equation}\label{OP1}
\Phi[g] = \Phi_{\rm ct} -T\int\frac{d^p\theta}{\Omega_p}d\tau dr\,(g_{rr}g_{\tau\tau})^{1/2}
= \Phi_{\rm ct} - \int_{r_{\rm min}}^\infty dr\,(g_{rr}g_{\tau\tau})^{1/2}\,.
\end{equation}
Varying the Lagrange multiplier term $-\kappa(\Phi[g]-\Phi_0)/T$ with respect to the metric gives a contribution to the stress tensor which is that of a relativistic string of tension $\kappa$ extended in the $\tau$ and $r$ directions and smeared over the angular directions. The Einstein equation is now:
\begin{equation}\label{Einstein}
{G^\tau}_\tau = {G^r}_r = \frac12p(p+1) - \frac{8\pi G_{\rm N}\kappa}{\Omega_pg_{\Omega\Omega}^{1/2}}\,, \qquad
{G^\theta}_\theta = \frac12p(p+1)\,.
\end{equation}
The first term on the two right-hand sides is the cosmological constant. On the right-hand side of ${G^\tau}_\tau$ and ${G^r}_r$, the string tension $\kappa$ is multiplied by the gravitational coupling and divided by the proper area of the $p$-sphere (since the string tension is spread out over that area).

It is well known (see \cite{Jacobson:2007tj} for a discussion) that, in the presence of spherical symmetry, if the stress tensor obeys ${T^\tau}_\tau = {T^r}_r$ and ${T^\tau}_r = {T^r}_\tau = 0$, then the solution to the Einstein equation can be written in the Schwarzschild form,
\begin{equation}\label{Smetric}
ds^2 = f(r)d\tau^2 + f(r)^{-1}dr^2 + r^2d\Omega_p^2\,.
\end{equation}
The function $f$ is easily found, and is a remarkably simple generalization of the AdS-Schwarz\-schild solution:
\begin{equation}\label{fdef}
f(r) = r^2 + 1  - \frac{16\pi G_{\rm N}\kappa}{p\Omega_pr^{p-2}} - \frac\mu{r^{p-1}}\,.
\end{equation}
As in the case of AdS-Schwarzschild (Section 2 above), $\mu$ is fixed by the requirement of smoothness of the metric at the horizon, which requires $f'(\rh)=4\pi T$, where the horizon radius $\rh$ is the largest root of $f$. The solution \eqref{Smetric}, \eqref{fdef} was found (for $p=2$) by Guendelman and Rabinowitz \cite{Guendelman:1991qb}, who termed it a ``hedgehog black hole".

At long distances, where the gravitational field is weak, the hedgehog black hole metric can be derived from Newtonian reasoning as follows. As usual in the weak field regime we have $f(r) = 1 + 2V(r)$, where $V$ is the Newtonian gravitational potential. Thanks to the spherical symmetry, the latter is given by
\begin{equation}
V(r) = -\frac{8\pi G_{\rm N}M(r)}{p\Omega_pr^{p-1}}\,,
\end{equation}
where $M(r)$ is the total mass contained within a ball of radius $r$. In the case of AdS-Schwarzschild, $M(r)$ includes a negative contribution proportional to $r^{p+1}$ from the vacuum energy, plus an arbitrary constant representing a mass located at the origin. In the present case there is also a contribution $\kappa r$ due to the string. The fact that this Newtonian reasoning gives the correct exact solution to the full Einstein equation can be traced to the facts that, for a metric of the form \eqref{fdef}, the Einstein tensor (with one raised index) is linear in $f$, and the stress tensor (with one raised index) is independent of $f$.\footnote{The Newtonian reasoning continues to give the correct solution if we also include a spherically symmetric electric field; we simply add the term $G_{\rm N}q^2/r^{2p-2}$ to the right-hand side of \eqref{fdef}, where $q$ is the electric charge of the hedgehog black hole.}

\FIGURE{
\epsfig{file=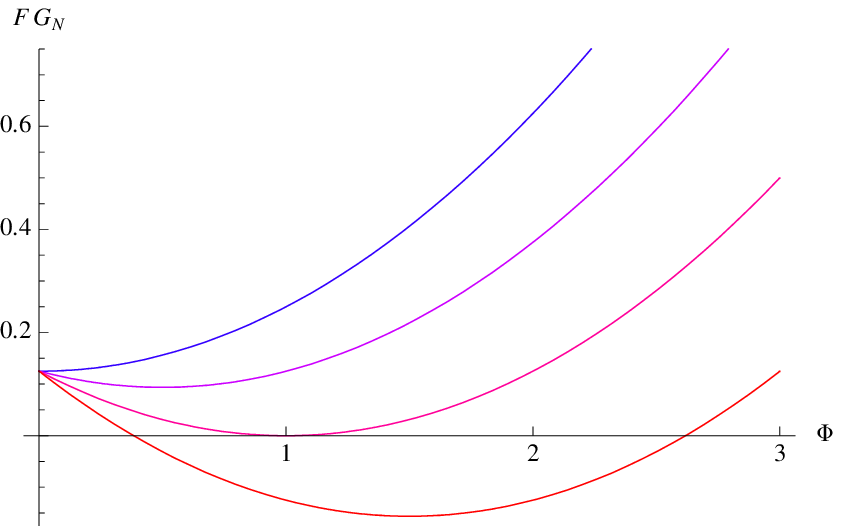,width=3 in}
\caption{\label{FpureS1}
$F$ versus $\Phi$ in three dimensions ($p=1$), at $4\pi T = 0,1,2,3$ (top to bottom). The minimum of each curve is a BTZ black hole (at $T=0$ the minimum is at $\Phi=0$ and represents the zero-mass black hole). The third curve is at the Hawking-Page temperature.
}}

We wish to compute, for the family of metrics obtained above, the relation between the order parameter $\Phi$ and the free energy $F$, which is $T$ times the (renormalized) Euclidean action. $\Phi$ is rather trivial to compute, and in fact doesn't depend at all the particular form of $f$; as in the case of AdS-Schwarzschild, we have
\begin{equation}\label{Phirh}
\Phi = \rh\,.
\end{equation}
The free energy can be obtained by directly computing the Euclidean action in an appropriate regularization and subtraction scheme (such as the one discussed in footnote \ref{regularization}). The calculation is straightforward but rather tedious. A simpler but less direct approach, which yields the same result, is to first find the relation between $\Phi$ and $\kappa$, and then use the relation
\begin{equation}
\frac{dF}{d\Phi} = \kappa\,
\end{equation}
to find $F$ as a function of $\Phi$. The relation between $\Phi$ and $\kappa$ can be found by combining the equations $f(\rh)=0$ and $f'(\rh)=4\pi T$ to eliminate $\mu$, and then using \eqref{Phirh}; the result is:
\begin{equation}
\frac{16\pi G_{\rm N}\kappa}{p\Omega_p} = (p+1)\Phi^p - 4\pi T\Phi^{p-1} + (p-1)\Phi^{p-2}\,,
\end{equation}
from which we obtain
\begin{equation}
\frac{16\pi G_{\rm N}F}{\Omega_p} = p\Phi^{p+1} - 4\pi T\Phi^p + p\Phi^{p-1}\,.
\end{equation}
The integration constant can be fixed, for $p>2$, by noting that the solutions go over in the limit $\Phi\to0$ to AdS${}_{p+2}$, which by convention has vanishing free energy. For $p\le2$, it is necessary to resort to the direct calculation of the Euclidean action.

In figure \ref{FpureS1} we plot $F$ versus $\Phi$ for $p=1$, and in figure \ref{kappaFpureS3} both $\kappa$ and $F$ for $p=3$, at various temperatures. All of the important thermodynamic features are clear from these figures: for $p=3$, the appearance of the large and small black holes at $T_{\rm crit}$, the instability of the small black hole and stability of the large one, and the fact that the large black hole becomes dominant at $T_{\rm HP}$; for $p=1$, the absence of a small black hole and the fact that the large black hole exists even at zero temperature (although it is not thermodynamically dominant there).

\FIGURE{
\epsfig{file=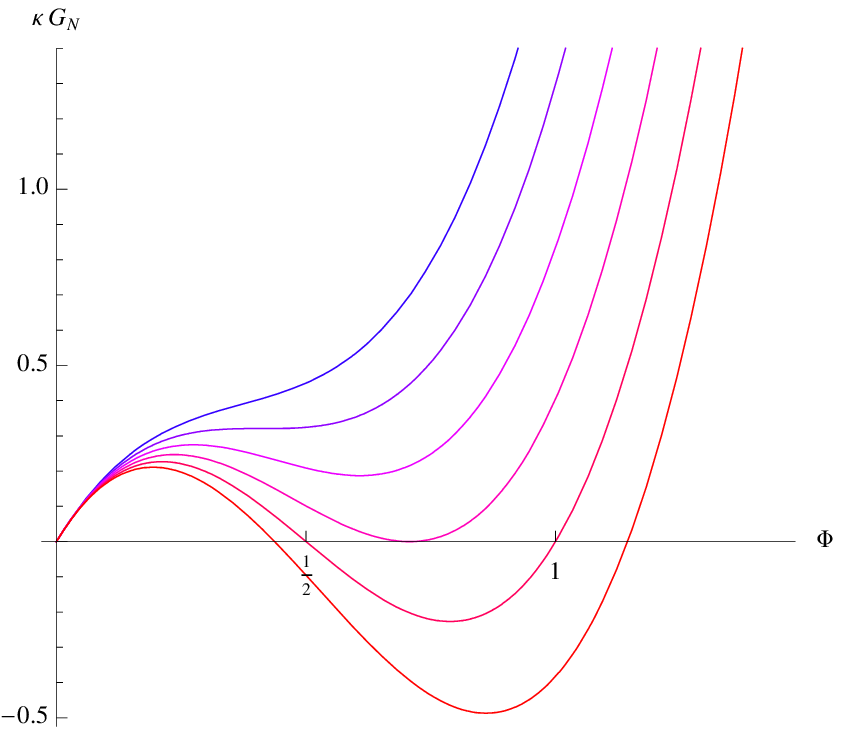,width=3.in}
\epsfig{file=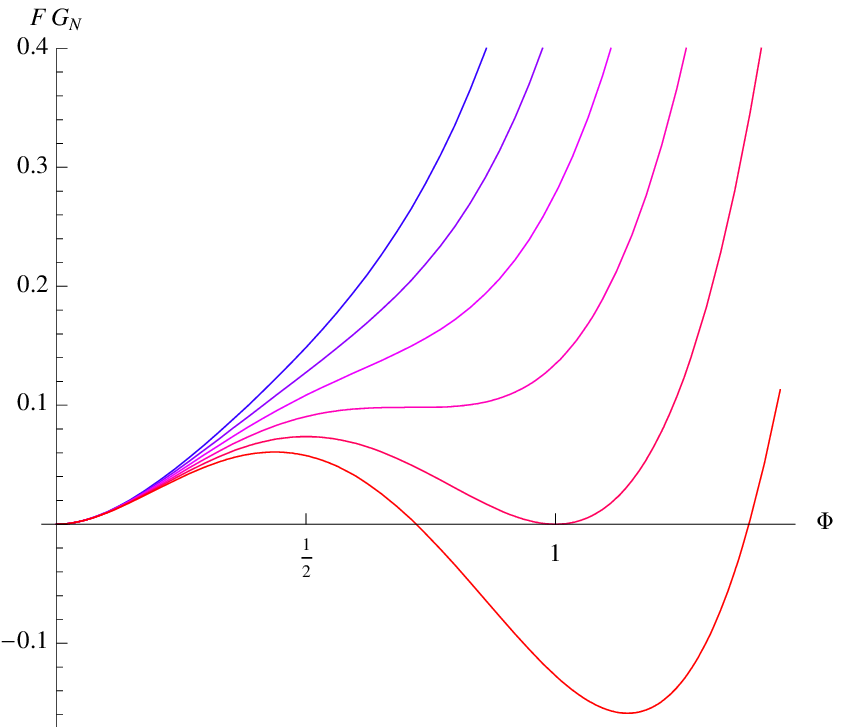,width=3.in}
\caption{\label{kappaFpureS3}
$\kappa$ and $F$ versus $\Phi$ in five dimensions ($p=3$), at $(2\pi T)^2 = 5,6,7,8,9,10$ (top to bottom). From the bottom, the second curve is at $T_{\rm HP}$, the third at $T_{\rm crit}$, and the fifth at $T_{\rm crit}'$ (where the curves' inflection points merge and disappear).
}}

\FIGURE{
\epsfig{file=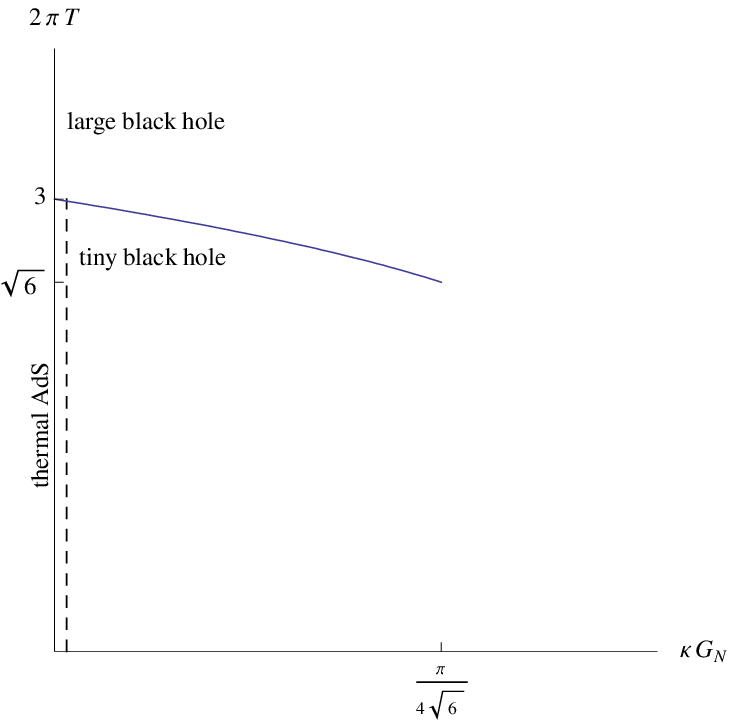,width=3.6in}
\caption{\label{phasepureS3}
Phase diagram in the $\kappa$--$T$ plane for the five-dimensional system ($p=3$), whose free energy diagram is shown in figure \ref{kappaFpureS3}. The dashed line represents a second-order transition at an infinitesimal value of $\kappa$ separating thermal AdS from a tiny black hole. The solid curve represents a first-order transition separating the tiny and large black hole phases. This is an extension of the Hawking-Page transition and ends at a critical point,.
}}

Let us focus on the case $p=3$, and ask where AdS${}_5$ belongs on the free energy diagram. (Strictly speaking, since we are dealing here with a toy model, this is an artificial question; however, let us pretend for the moment that we are dealing with a real string theory.) As mentioned in the previous section, for AdS we have $\Phi = -2\pi T\lambda^{-1/2}\ln N$, whose value at large $N$ depends on how one takes the limit. The conventional 't Hooft limit ($N\to\infty$, $\lambda$ fixed) yields $\Phi = -\infty$. The free energy diagram then seems to suffer from a horizontal discontinuity, with no solutions connecting AdS${}_5$ at $\Phi=-\infty$ to the hedgehog black holes at $\Phi>0$ (the hedgehog black holes with $\Phi\le0$ are nakedly singular). However, we should recall that in the 't Hooft limit the string length $\lambda^{-1/2}$ remains fixed. In particular, hedgehogs with $\Phi$ of order $\lambda^{-1/2}$ will experience large stringy corrections near the horizon (recall that $\rh=\Phi$). In other contexts, such corrections are known to smooth out topology-changing transitions; one may therefore expect that, once they are taken into account here, the hedgehogs will extend all the way to $\Phi=-\infty$ and connect smoothly to AdS. Less speculatively, one can choose to take the large $N$ limit in such a way that the stringy corrections disappear at the same time as the quantum corrections, for example by keeping $g_{\rm YM}^2$ fixed. This puts AdS at $\Phi=0$, and gives a continuous free energy diagram. This is the point of view we will adopt in the rest of the paper.

If we consider $\kappa$ to be a chemical potential rather than a Lagrange multiplier, in other words if we change the action by $-\kappa\Phi[g]/T$ rather than $-\kappa(\Phi[g]-\Phi_0)/T$, then we can use the above free energy diagrams to derive a phase diagram in the $\kappa$--$T$ plane. The effect of the chemical potential is to add a density of external ``quarks", which source strings that extend into the bulk and end on the horizon. Figure \ref{phasepureS3} shows the phase diagram for the five-dimensional ($p=3$) case. To understand this diagram, let us work at fixed temperature and study the effect of varying $\kappa$. Above the Hawking-Page temperature, the chemical potential has no qualitative effect, since the system is already in the large black hole state (it does make the horizon larger---the strings effectively pull on it). However, below the Hawking-Page temperature, where in the absence of a chemical potential there is no black hole, the strings force a horizon to open. This can be seen from the free energy diagram, which is quadratic in the neighborhood of the origin. Adding a chemical potential creates a local minimum at $\Phi\approx 4G_{\rm N}\kappa/(3\pi)$, which, for sufficiently small $\kappa$, is the global mininum. There is thus a second-order transition at $\kappa=0$, from the thermal AdS state to a ``tiny black hole" (not to be confused with the small black hole, which is thermodynamically unstable and therefore never dominant). For temperatures between $T_{\rm HP}$ and $T_{\rm crit}'=6^{1/2}/(2\pi)$, there is a further, first-order transition between the tiny and the large black holes, at
\begin{equation}
\kappa = \frac{\pi^2T}{4G_{\rm N}}\left(1-\left(\frac{2\pi T}3\right)^2\right).
\end{equation}
As we lower the temperature past $T_{\rm crit}'$, the tiny and large black hole saddle points merge, and the phase transition curve separating them ends at a critical point.

The same calculation can easily be done in the Poincar\'e patch of AdS, where the boundary ``gauge theory" lives on $\R^p$ rather than $S^p$. The resulting hedgehog black brane solution is
\begin{equation}
ds^2 = f(r)d\tau^2 + f(r)^{-1}dr^2 + r^2dx_p^2, \qquad
f(r) = r^2 - \frac{16\pi G_{\rm N}\kappa}{pr^{p-2}} - \frac\mu{r^{p-1}}\,.
\end{equation}
The effective potential for $\Phi$ is
\begin{equation}
16\pi G_{\rm N}F = p\Phi^{p+1} - 4\pi T\Phi^p\,.
\end{equation}
(Here $\kappa$ and $F$ are the string tension and free energy per unit volume on the boundary.) The Hawking-Page transition in this system occurs at zero temperature and is second-order.

\section{Supergravity}

\subsection{Gauge theory on $S^3$}

In this section, we extend the calculation of the previous section to type IIB supergravity on AdS${}_5\times S^5$. The boundary conditions for the problem have an $SO(2)\times SO(4)\times SO(6)$ symmetry. For the time being, we will work within an ansatz for the supergravity fields that respects this symmetry; below, we will consider the issue of spontaneous symmetry breaking (particularly of the $SO(6)$ factor). The only supergravity fields allowed by these symmetries are the following: the metric, which must be of the form
\begin{equation}
ds^2 = 
g_{\tau\tau}(r)d\tau^2 + g_{rr}(r)dr^2 + e^{2\sigma(r)}d\Omega_3^2 + e^{2\ts(r)}d\Omega_5^2\,,
\end{equation}
(this is the string-frame metric); the dilaton $\Phi(r)$ (not to be confused with our order parameter); the Ramond-Ramond scalar, which is not sourced by the Lagrange multiplier term and can consistently be set to zero; and the five-form field strength, which, taking into account the flux quantization on the $S^5$, must be of the form
\begin{equation}
F_5 = 16\pi N\alpha'^2\left(d^5\tilde\theta + *d^5\tilde\theta\right),
\end{equation}
where $d^5\tilde\theta$ is the volume form on the unit $S^5$. The five-form thus does not carry any independent dynamical degrees of freedom, and we have in total five degrees of freedom minus one gauge degree of freedom. Within this ansatz, the supergravity equations of motion can be derived from the following action:
\begin{multline}\label{action}
S = - \frac{N^2}{4T}\int dr\,(g_{\tau\tau}g_{rr})^{1/2}e^{-2\phi} \\
\times\left(
R_{(2)} + 
g^{rr}\left(4(\partial_r\phi)^2 - 3(\partial_r\sigma)^2 - 5(\partial_r\ts)^2\right)
+ 6e^{-2\sigma} + 20e^{-2\ts} - 8e^{2\phi+3\sigma-5\ts}
\right),
\end{multline}
where $R_{(2)}$ is the Ricci scalar of the two-dimensional metric $g_{\tau\tau}d\tau^2+g_{rr}dr^2$, and $\phi \equiv \Phi - \frac32\sigma - \frac52\ts$ is the two-dimensional dilaton. (When directly computing the on-shell value of the Euclidean action, one must also take into account a counterterm and boundary term.) The order parameter is, as in \eqref{OP1},
\begin{equation}
\Phi[g] = -\int dr\,(g_{rr}g_{\tau\tau})^{1/2}\,.
\end{equation}
When added to the action, the Lagrange multiplier term $-\kappa(\Phi[g] - \Phi_0)/T$ sources the metric, which sources $\phi$, which in turn sources $\sigma$ and $\ts$. Therefore all the above fields are active in the solutions.

After gauge fixing, the equations of motion derived from the action $S - \kappa(\Phi[g]-\Phi_0)/T$ reduce to a system of four coupled ordinary differential equations. The boundary conditions on the horizon are determined by the requirement of smoothness of the metric and scalar fields. On the outer boundary, at $r=\infty$, we have asymptotically AdS${}_5^T\times S^5$ boundary conditions (with Dirichlet boundary conditions on the dilaton); in practice these boundary conditions were imposed by matching onto an asymptotic solution at a finite value of $r$. The resulting boundary value problem was solved numerically using a shooting method, implemented in \emph{Mathematica} using a combination of the NDSolve and FindRoot routines.

\FIGURE{\epsfig{file=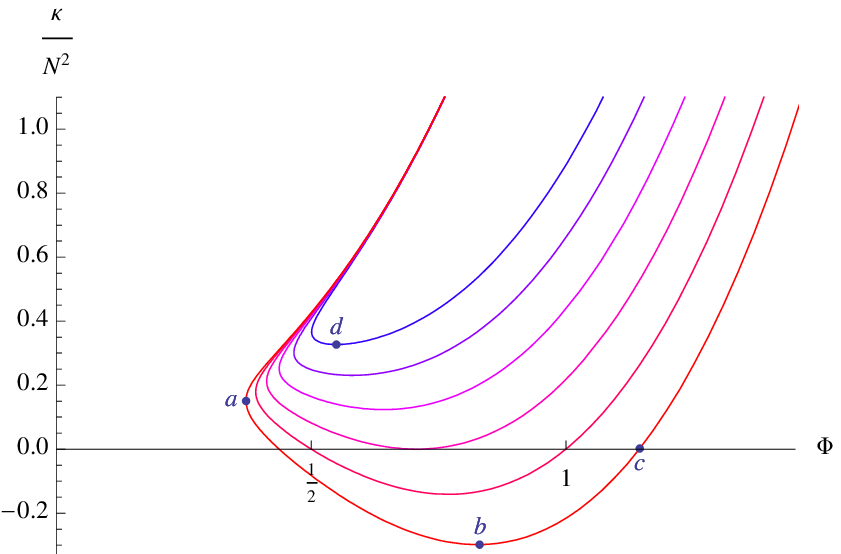,width=4.4in}\caption{\label{kappaS3}
$\kappa$ versus $\Phi$ for the same set of temperatures as shown in figure \ref{kappaFpureS3}, namely (from top to bottom) $(2\pi T)^2 = 5,6,7,8,9,10$. $2\pi T=8^{1/2}$ is the critical temperature, while $2\pi T=3$ is the Hawking-Page temperature. (The curves for some higher temperatures are shown in figure \ref{kappaR3}.) The solutions corresponding to the points labelled $a$, $b$, $c$, $d$ are plotted in figure \ref{solutionplotsS3}.
}}

\FIGURE{\epsfig{file=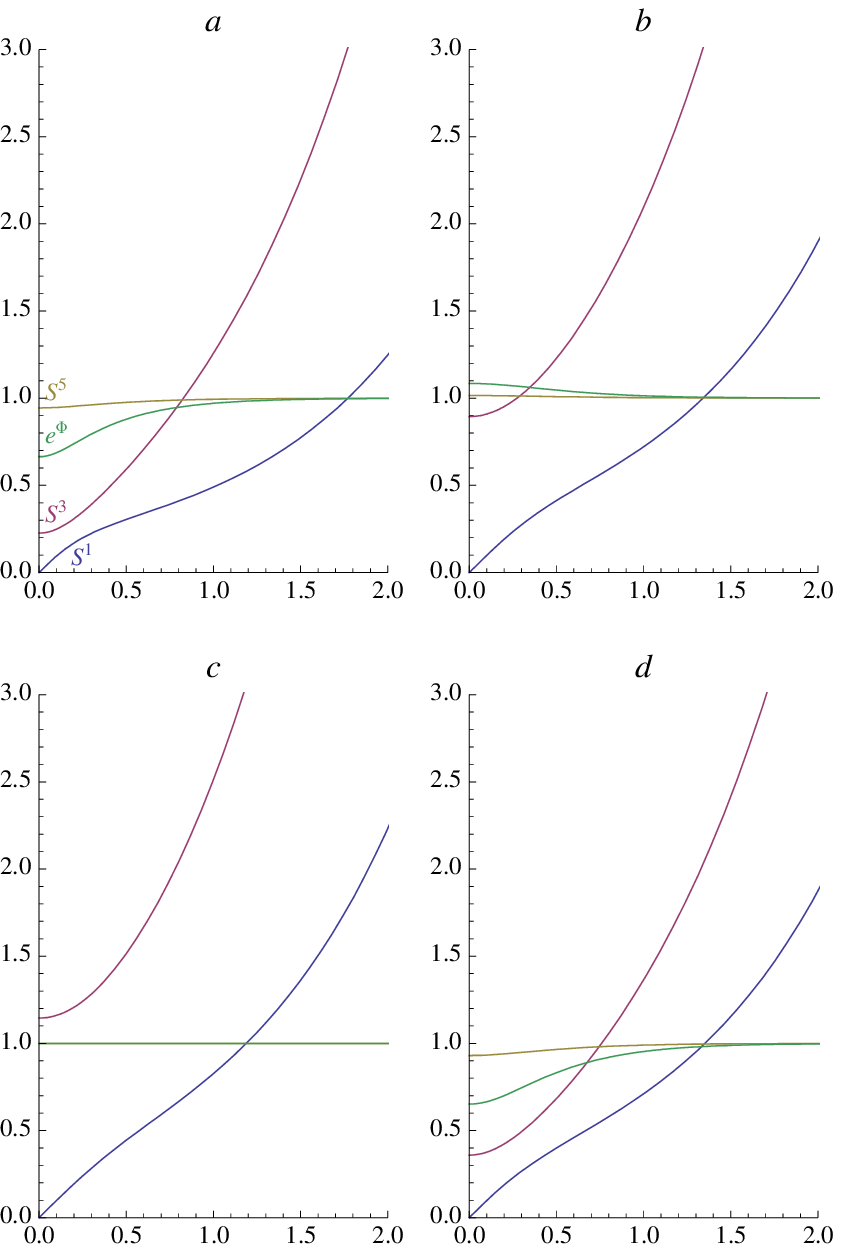,width=4.2in}\caption{\label{solutionplotsS3}
A selection of supergravity hedgehog black hole solutions, corresponding to the labelled points in figure \ref{kappaS3}. $a$: $(2\pi T, \Phi, \kappa/N^2, F/N^2) = (10^{1/2}, 0.37, 0.15, 0.04)$; $b$: $(10^{1/2}, 0.83, -0.30, -0.04)$; $c$: $(10^{1/2}, 1.14, 0, -0.10)$ (this is the conventional large black hole at this temperature); $d$: $(5^{1/2}, 0.55, 0.33, 0.09)$. The horizontal axis is proper radial distance from the horizon. The curves marked $S^{1,3,5}$ represent the proper radii of the respective spheres, and the one marked $e^\Phi$ represents the local string coupling (relative to its asymptotic value $g_{\rm s}$). In case $c$ (the conventional black hole) both the dilaton and $S^5$ radius are constant.
}}

The results are shown in figure \ref{kappaS3} in the form of plots of $\kappa$ versus $\Phi$, at the same six temperatures as in figure \ref{kappaFpureS3}. The points where the curves intersect the horizontal axis represent conventional black holes. Above the critical temperature $T_{\rm crit}=8^{1/2}/(2\pi)$ (bottom two curves), the curves cross the axis twice, representing the large and small black holes. At the critical temperature (third from bottom) it is tangent to the axis, while below the critical temperature (top three curves) they do not intersect it at all. The major new feature compared to the same plot for pure gravity (left side of figure \ref{kappaFpureS3}) is that, whereas the latter curves go through the origin, the curves for the supergravity system seem to be ``repelled" by the origin. At each temperature there is a positive minimum value of $\Phi$ below which there are no solutions, and above which there are two. Thus it would seem that we have failed in our attempt to fix the value of the order parameter, at least to certain values. We will discuss a possible resolution to this issue in the next subsection. It is also interesting to note that the upper branches of the curves quickly merge with each other. Inspection of the corresponding solutions shows that those with the same $\Phi$ but different $T$ on that branch are essentially identical outside of a neighborhood of the horizon, and that neighborhood becomes smaller and smaller as $\Phi$ increases. A few examples of solutions are plotted in figure \ref{solutionplotsS3}.

\FIGURE{\epsfig{file=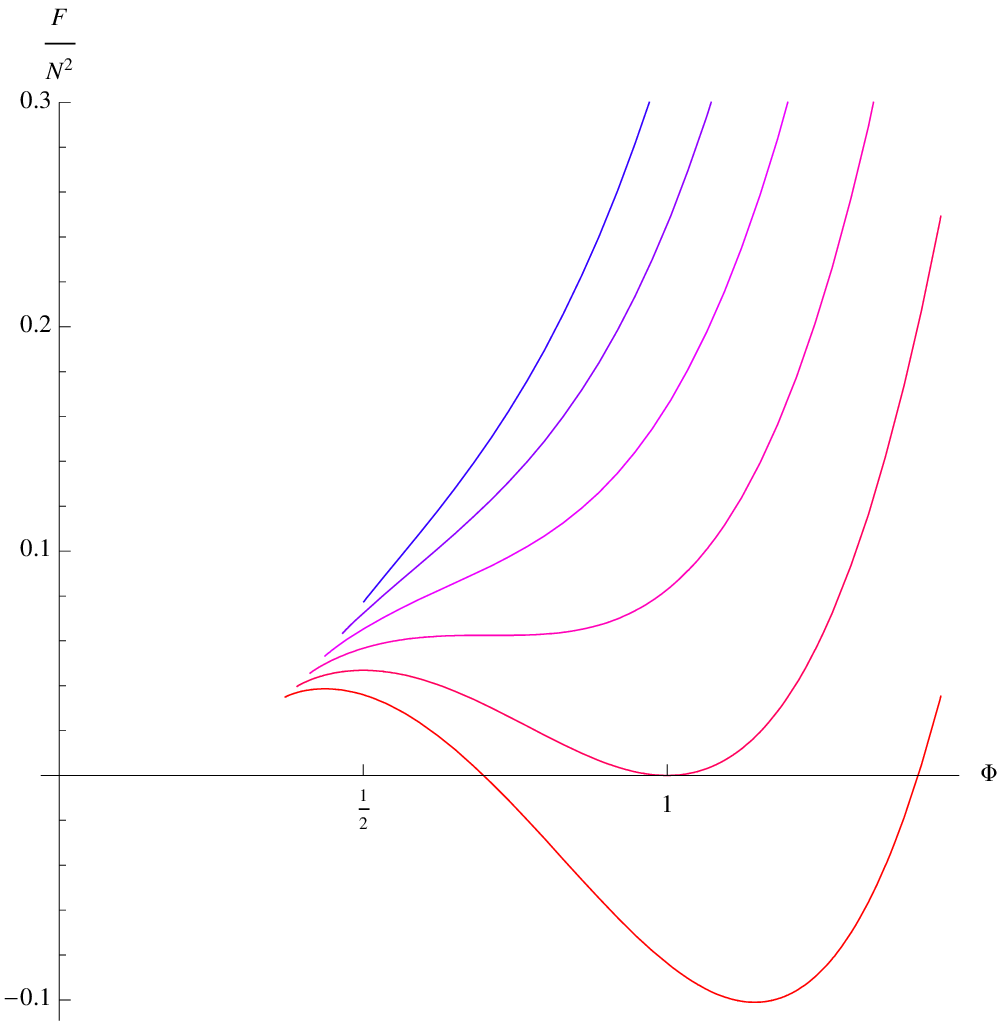,width=5in}\caption{\label{FS3}
Free energy diagram for the same set of temperatures as in previous figures. Only the lower branch of solutions is plotted, since they obviously have lower free enegy at a given value of $\Phi$ than the solutions on the upper branch.
}}

The free energy can be computed from the curves of figure \ref{kappaS3} by using the relation $dF/d\Phi=\kappa$.\footnote{
The integration constants were fixed as follows. First, the large black hole at the Hawking-Page temperature ($2\pi T=3$, $\Phi=1$) has vanishing action, fixing the integration constant for that temperature. The other temperatures can then be fixed by matching in the region where the curves for the different temperatures merge, namely along the upper branch of solutions in the $\Phi$--$\kappa$ plane.
} It is plotted in figure \ref{FS3}. For each temperature, we have only plotted the branch of solutions with the lower value of $\kappa$, since these clearly also have the lower value of $F$. The curves are qualitatively similar to the case of pure gravity (right side of figure \ref{kappaFpureS3}), and again exhibit the major thermodynamic features of the system: the appearance of the large and small black holes at $T_{\rm crit}$, the thermodynamic stability of the large black hole and instability of the small one, and the fact that the large black hole becomes dominant at $T_{\rm HP}$. The major difference with the case of pure gravity is, again, that each curve ends at a minimum value of $\Phi$.

\FIGURE{\epsfig{file=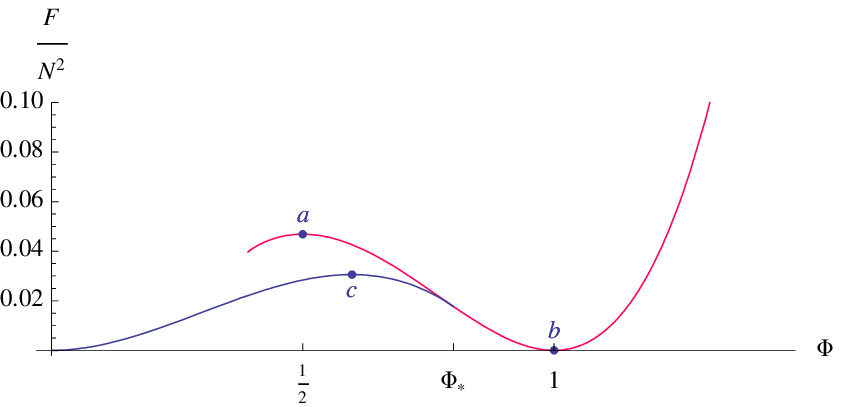,width=5in}\caption{\label{sketch}
Conjectural sketch of the complete free energy diagram, at the representative temperature $T_{\rm HP}$, when the possibility of spontaneous $SO(6)$ breaking is included. The upper curve is the same one shown in the previous figure, and represents $SO(6)$-preserving hedgehogs. The points marked $a$ and $b$ are the five-dimensional small and large black holes respectively. The lower curve is a conjectural sketch of a new branch of $SO(6)$-breaking hedgehogs, which range from $\Phi=0$ (thermal AdS) to $\Phi=\Phi_*$, where they connect continuously onto the upper curve. The maximum of this new branch ($c$) is a conventional black hole (since it has $\kappa=0$), and is presumably the ten-dimensional black hole, which is thermodynamically unstable but otherwise stable. The value of $\Phi_*$ could be determined by a linearized perturbation analysis of the $SO(6)$-preserving hedgehogs; the rest of the curve would require a difficult computation (a warm-up problem would be to find the conventional ten-dimensional black hole solution $c$). Somewhere between the point $c$ and $\Phi_*$, the new branch of hedgehogs would undergo a topology-changing transition, from ten-dimensional black holes localized on the $S^5$ to five-dimensional black holes that are non-uniform over the $S^5$.
}}

Our discussion so far has neglected the possibility of spontaneous breaking of the system's $SO(2)\times SO(4)\times SO(6)$ symmetry. Searching for solutions to the full ten-dimensional supergravity equations of motion, without imposing those symmetries, would present a challenging numerical problem, and no attempt was made to do so here. (Aside from the increase in dimensionality, a complication that arises when the symmetries are not imposed is that, as mentioned in Section 3, $\Phi[g]$ becomes a non-local functional. To include the effect of the Lagrange multiplier, it is therefore necessary to solve simultaneously the string's equations of motion and the gravitational ones. Up to now we have only had to solve the gravitational equations, since the symmetries restrict the string world-sheet to lie in the $\tau$--$r$ plane at fixed $\theta$ and $\tilde\theta$.) However, some information can be obtained from the known properties of the conventional black holes in this system, which were reviewed at the end of Section  2. In particular, the large black hole is always locally stable, whereas the small black hole suffers from a Gregory-Laflamme instability on the $S^5$ at temperatures above $T_{\rm GL}\approx3.20/(2\pi)$ \cite{Hubeny:2002xn}. While we have not done the linearized perturbation analysis for our hedgehog solutions (it should not be too difficult in principle), the simplest picture consistent with the above facts is the following: at each temperature, there is a critical value $\Phi_*(T)$ such that, for $\Phi>\Phi_*(T)$ the hedgehog is stable, and for $\Phi<\Phi_*(T)$ it is unstable against $SO(6)$-breaking perturbations; furthermore, for $T$ in the range $T_{\rm crit}<T<T_{\rm GL}$, $\Phi_*(T)<\Phi_{\text{small black hole}}$, while for $T>T_{\rm GL}$, $\Phi_{\text{small black hole}}<\Phi_*(T)<\Phi_{\text{large black hole}}$. For $\Phi<\Phi_*$, there should exist a branch of $SO(6)$-breaking hedgehogs with free energies less than the $SO(6)$-preserving ones plotted in figure \ref{FS3}, which continuously connects onto those solutions at $\Phi_*$. This new branch would include both ten-dimensional black holes localized on the $S^5$, and five-dimensional ones that are non-uniform on the $S^5$. The ten-dimensional conventional black hole should correspond to a local maximum on this new branch (maximum because that black hole is thermodynamically unstable). Finally, we conjecture that that branch extends all the way down to $\Phi=0$, where it continuously connects onto thermal AdS, as occurred for pure gravity; the $SO(6)$ symmety is restored as the tiny 10-dimensional black hole disappears.
Figure \ref{sketch} shows a sketch of this conjectured branch of $SO(6)$-breaking hedgehogs, superimposed on the $SO(6)$-preserving free energy diagram at the Hawking-Page temperature.

\FIGURE{\epsfig{file=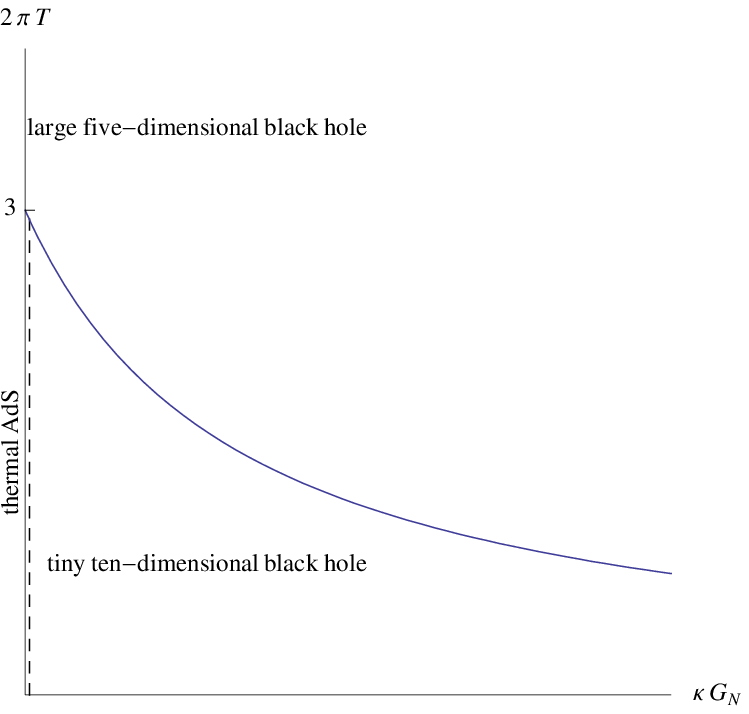,width=3.7in}\caption{\label{phaseS3}
Conjectural sketch of the phase diagram in the presence of a chemical potential $\kappa$ for the PML.
}}

If the free energy diagram sketched in the previous paragraph and in figure \ref{sketch} is correct, then we can derive from it the phase diagram for the system when we consider $\kappa$ to be a chemical potential rather than a Lagrange multiplier, that is, when we add $-\kappa\Phi[g]/T$ rather than $-\kappa(\Phi[g]-\Phi_0)/T$ to the action. Such a chemical potential for the Polyakov-Maldacena loop (PML) corresponds roughly to a density $\kappa$ of external quarks in the gauge theory. As in pure five-dimensional gravity analyzed at the end of the previous section, above the Hawking-Page temperature turning on $\kappa$ has no qualitative effect. Below that temperature, however, there is a second-order transition at $\kappa=0$ to a phase with a tiny ten-dimensional hedgehog black hole. This black hole is smaller than the conventional ten-dimensional black hole, and exists only by virtue of the strings sourced by the external quarks, whose tension holds the horizon open. As $\kappa$ is increased, at some point there is a first-order transition in which the system flips to the large five-dimensional black hole. Unlike in the case of pure gravity (figure \ref{phasepureS3}), the curve separating the tiny and large black hole phases cannot end, since the two phases have different symmetries ($SO(6)$ is broken in one and not the other). This phase diagram is sketched in figure \ref{phaseS3}. Of course, this picture is only the simplest one compatible with what we know at present, and the true phase diagram could be more complicated.

\subsection{Gauge theory on $\R^3$}

\FIGURE{
\epsfig{file=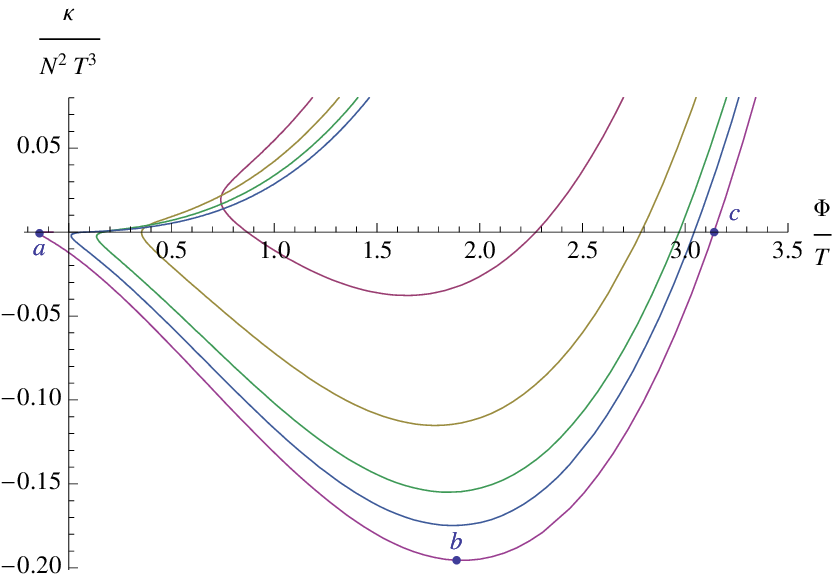,width=3.4in}
\epsfig{file=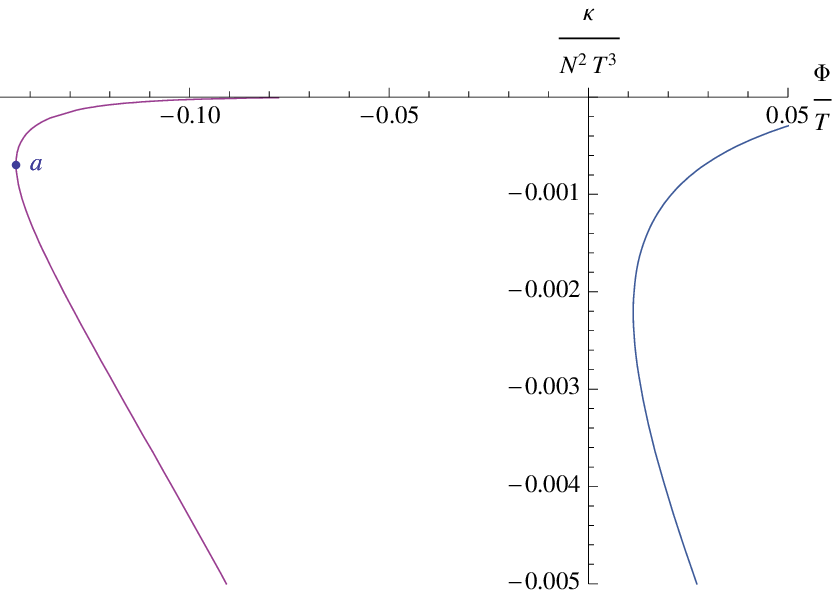,width=2.6in}
\caption{\label{kappaR3}
Left: $\kappa$ versus $\Phi$ for (top to bottom) $(2\pi T)^2 = 10,20,40,80,\infty$. The bottom curve is the result for the gauge theory on $\R^3$. Note that each curve---including the one for $\R^3$---goes to a minimum value of $\Phi$. Here $\kappa$ is the string tension per unit volume on the boundary; $\kappa_{\rm here} = \kappa_{\rm above}/\Omega_3$. Right: The left figure blown up near the origin, showing the curves for $(2\pi T)^2=\infty$ (left) and $80$ (right). The curve for $T=\infty$ presumably ends at the origin, representing AdS, although this region is difficult to access computationally. The solutions corresponding to the points $a$, $b$, $c$ are plotted in figure \ref{solutionplotsR3}.
}}

\FIGURE{
\epsfig{file=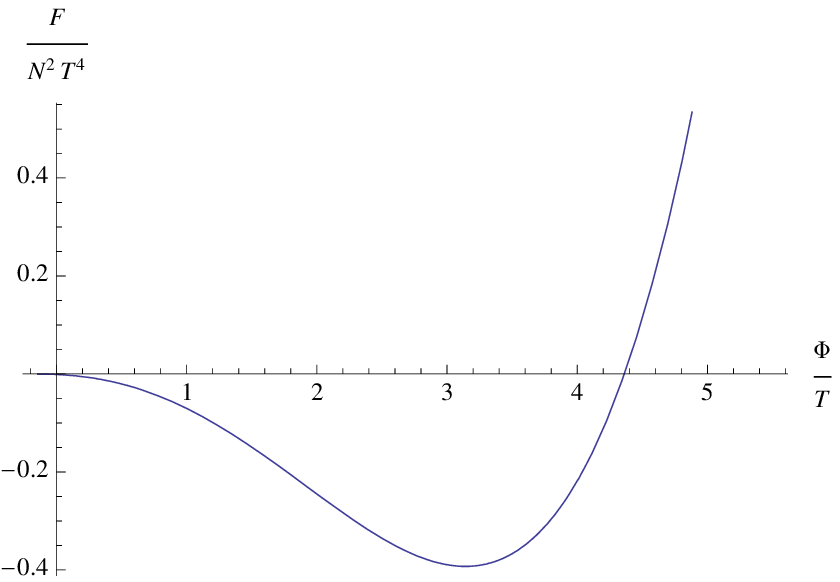,width=3.4in}
\epsfig{file=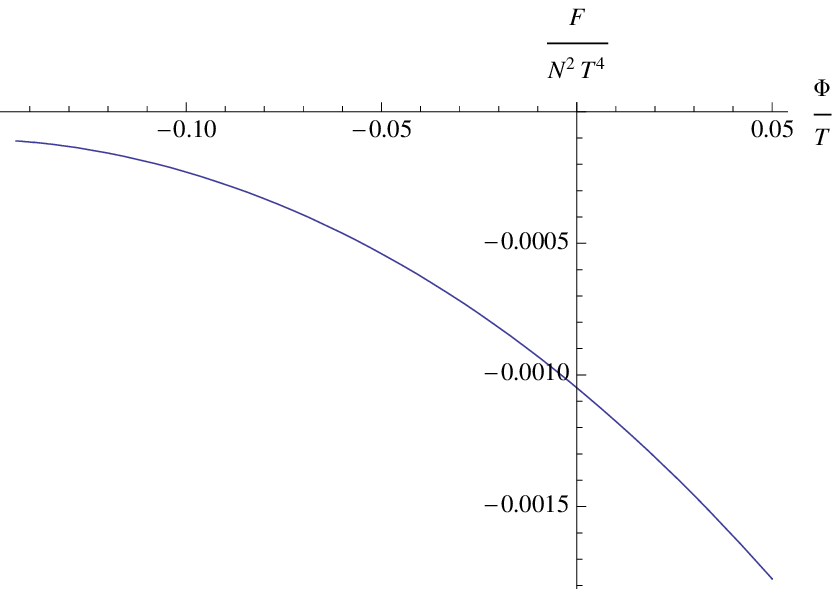,width=2.6in}
\caption{\label{FR3}
Left: Free energy diagram for the gauge theory on $\R^3$. Here $F$ is the free energy per unit volume on the boundary. Right: left figure blown up near the origin.
}}

\FIGURE{\epsfig{file=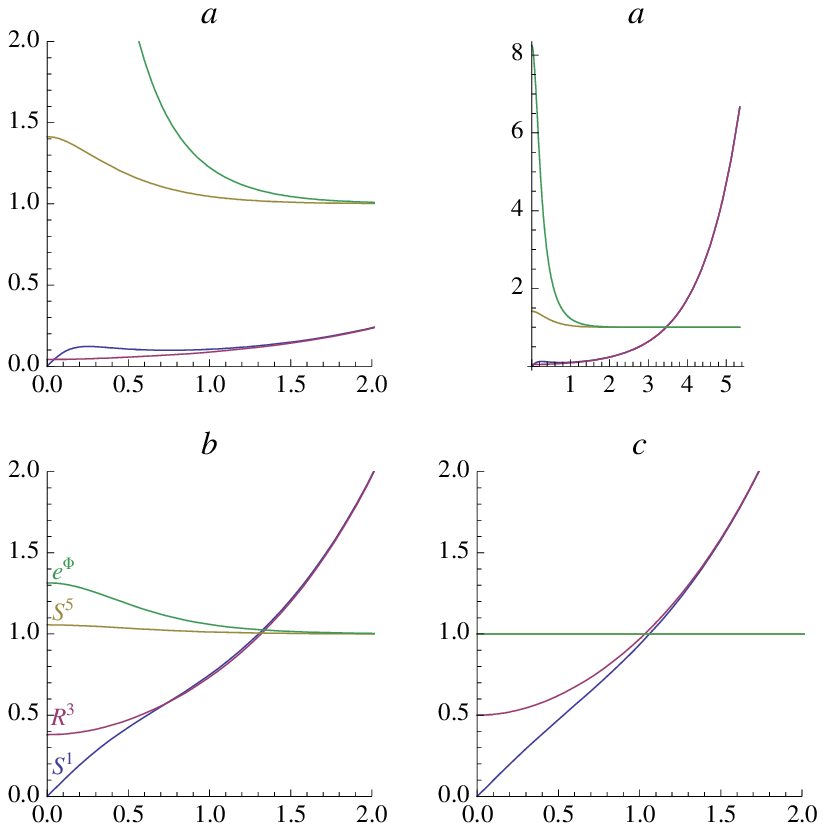,width=5in}\caption{\label{solutionplotsR3}
A selection of supergravity hedgehog black brane solutions, corresponding to the labelled points in figure \ref{kappaR3}. $a$ (shown twice for clarity): $(\Phi/T,\kappa/N^2T^3,F/N^2T^4) = (-0.14, -0.0007, -0.0001)$; $b$: $(1.9, -0.20, -0.22)$; $c$: $(\pi, 0, -0.39)$ (this is the conventional black brane). The horizontal axis is proper radial distance from the horizon. The curves marked $S^1$ and $S^5$ represent the proper radii of the respective spheres, the one marked $R^3$ represents (the square root of) the warp factor for the $\R^3$ part of the metric, and the one marked $e^\Phi$ represents the local string coupling (relative to its asymptotic value $g_{\rm s}$). In case $c$ both the dilaton and $S^5$ radius are constant.
}}

On $\R^3$, unlike on $S^3$, different temperatures are related by conformal transformations, and therefore exhibit the same physics. There is therefore only one free energy diagram to compute, which is $F/T^4$ versus $\Phi/T$. The only effect on the action \eqref{action} of going to $\R^3$ is to remove the term $6e^{-2\sigma}$ from the potential. The theory on $\R^3$ can be considered as the $T\to\infty$ limit of the theory on $S^3$. In figure \ref{kappaR3} we plot $\kappa/T^3$ versus $\Phi/T$ on $S^3$ for a range of high temperatures, together with the result on $\R^3$. Three examples of hedgehog black brane solutions are plotted in figure \ref{solutionplotsR3}. Figure \ref{FR3} shows the free energy diagram on $\R^3$, obtained by integrating the (lowest) $\Phi$--$\kappa$ curve in figure \ref{kappaR3}.

\newpage

\acknowledgments

I would like to thank the following individuals for many useful ideas and discussions: O. Aharony, R. Bousso, H. Elvang, S. Hartnoll, T. Jacobson, A. Lawrence, S. Minwalla, S. Mukhi, S. Shenker, B. Tekin, M. \"Unsal, M. Van Raamsdonk, S. Wadia, T. Wiseman, and L. Yaffe. I would like to thank T. Wiseman in particular for crucial help with the numerics, and him and M. \"Unsal for helpful suggestions on the manuscript. I would also like to thank the following institutions for their hospitality while this work was being done: the Aspen Center for Physics, the Kavli Institute for Theoretical Physics, the MIT Center for Theoretical Physics, Imperial College London, the Tata Institute of Fundamental Research, and the Isaac Newton Institute. I am supported by the Stanford Institute for Theoretical Physics and by NSF grant PHY 9870115.

\bibliography{ref}
\bibliographystyle{JHEP}

\end{document}